\documentclass[12pt,preprint]{aastex}

\slugcomment{To appear in Astronomical Journal}

\shorttitle{Photometry of $h$ \& $\chi$ Persei}
\shortauthors{Keller et al.}
\begin{document}
 
\title{UBVI and H$\alpha$ Photometry of the $h$ \& $\chi$ Persei cluster}


\author{Stefan C. Keller,}
\affil{Institute of Geophysics and Planetary Physics, \\
Lawerence Livermore National Laboratory, \\ 7000 East Ave., Livermore, CA 94550}
\email{skeller@igpp.ucllnl.org}

\author{Eva K. Grebel}
\affil{Max Planck Institute for Astronomy, \\ K\"onigstuhl 17, D-69117 Heidelberg, Germany}

\author{Grant J. Miller}
\affil{Southwestern College, \\
 900 Otay Lakes Road, Chula Vista, CA 91910}

\and

\author{Kenneth M. Yoss}
\affil{Department of Astronomy, \\
 University of Illinois at Urbana-Champaign, \\
1002 West Green Street, Urbana, IL 61801}

\begin{abstract} 
$UBVI$ and H$\alpha$ photometry is presented for 17319 stars in
vicinity of the young double cluster $h$ \& $\chi$ Persei. Our
photometry extends over a 37\arcmin\ $\times$ 1\arcdeg\ field centered
on the association. We construct reddening contours within the imaged
field. We find that the two clusters share a common distance modulus
of 11.75$\pm$0.05 and ages of log age(yr) = 7.1$\pm$0.1. From the
$V$$-$H$\alpha$ colour, a measure of the H$\alpha$ emission strength,
we conduct a survey for emission line objects within the
association. We detect a sample of 33 Be stars, 8 of which are new
detections. We present a scenario of evolutionary enhancement of the
Be phenomenon to account for the peak in Be fraction towards the top
of the main-sequence in the population of $h$ \& $\chi$ Persei and
similar young clusters.
\end{abstract}
 
 
\section{Introduction}

The double cluster $h$ \& $\chi$ Persei is one of the richest young
galactic open clusters. The two clusters, $h$ (NGC 869) and $\chi$
(NGC 884) Persei form the nuclei of the broader Per OB1 association
(Morgan 1953). The region has been the focus of numerous
studies. An extensive photographic study was presented by Oosterhoff
\cite{oos37}; further photographic studies were made by Moffat \&
Vogt \cite{mof74}. Photometric studies were made by Johnson \&
Morgan \cite{joh55}, Wildey \cite{wid64}, Schild \cite{sch65} in
Johnson $UBV$ and by Crawford et al.\ \cite {cra70}, Waelkens et
al.\ \cite{wae90}, Fabregat et al.\ \cite{fab96} and Marco \&
Bernabeu \cite{mar01} in Str\"omgren $uvby\beta$. Infra-red
photometry is presented by Tapia et al.\ \cite{tap84}. A sample of
proper motions is given in the central regions of $h$ \& $\chi$
Per. by Muminov \cite{mum83}.

Given their relative richness and young age, these
clusters form a useful comparison with stellar evolutionary models for
massive stars. With MS populations up to $\sim$15M$_{\odot}$ the
clusters are analogous to the young populous clusters of the
Magellanic Clouds. Within the literature, however, there is no
convergence on the fundamental parameters of the clusters, such as
their distances and ages. Marco \& Bernabeu \cite{mar01} present
distance moduli (dm) of 11.66$\pm$0.20 and 11.56$\pm$0.20 for $h$ \&
$\chi$ Per. respectively. This is consistent with a common distance to
the two clusters such as derived by Crawford et al.\ \cite {cra70}:
dm=11.4$\pm$0.4 and Balona \& Shobbrook \cite{bal84}:
dm=11.16. Tapia et al.\ \cite{tap84} and Schild \cite{sch65} both
find moduli of 11.7 and 12.0 for $h$ \& $\chi$ Per. respectively.

The age structure of the association has been discussed by Wildey
\cite{wid64} and Schild \cite{sch67}. Wildey concludes that a
significant age spread is present within the larger Per OB1
association with three notable epochs. Schild elaborates upon this,
finding $h$ to be older than $\chi$ Per. and discerns a third
spatially diffuse population of yet younger age in the surrounding
field. The present work revisits these issues in the light of our
CCD-based photometry and the results of modern stellar evolutionary
models.

Previous CCD based studies have remained limited to the central-most
regions of the clusters. It is our goal in the present work to
provide CCD based photometry over an area broad enough to encompass
the extent of the clusters. With this photometric basis we determine
the fundamental parameters of the clusters: distance modulus,
reddening (Section \ref{reddening}) and age (Section \ref{secage}).

The presence of a large population of Be stars within the clusters has
been noted by many authors (see e.g.\ Slettebak 1985 and Maeder et
al.\ 1999). A Be star is a B-type star which has at some stage
shown emission in H Balmer lines. This emission is understood to arise
from a expelled circumstellar disk of material in the equatorial plane
of the star. The mechanism for expulsion of this material is a topic
of debate but is likely related to the generally rapid photospheric
rotational velocity of the underlying star. 

The means of Be star detection in these previous works has been
through grism or long-slit spectroscopy. Such techniques, which have
largely focused on the nuclei of the two clusters, do not provide a
sample with uniform completeness, especially for V$\geq$10. Our study
uses the $V$$-$H$\alpha$ colour index as a measure of the H$\alpha$
emission strength to distinguish the Be star population in a uniform
manner throughout the observed field. The fraction of Be stars to
total B stars within these clusters is a useful comparison with the Be
star fractions observed within Magellanic Cloud clusters of similar
ages (Keller et al.\ 2000). A recurring feature of the Be star
population within young clusters is the peak in Be fraction towards
the terminus of the main sequence. In Section \ref{secbe} we discuss
this feature and its implications for evolutionary modification of the
Be phenomenon.

\section{Observations and Data Reduction}

\begin{figure}
\plotone{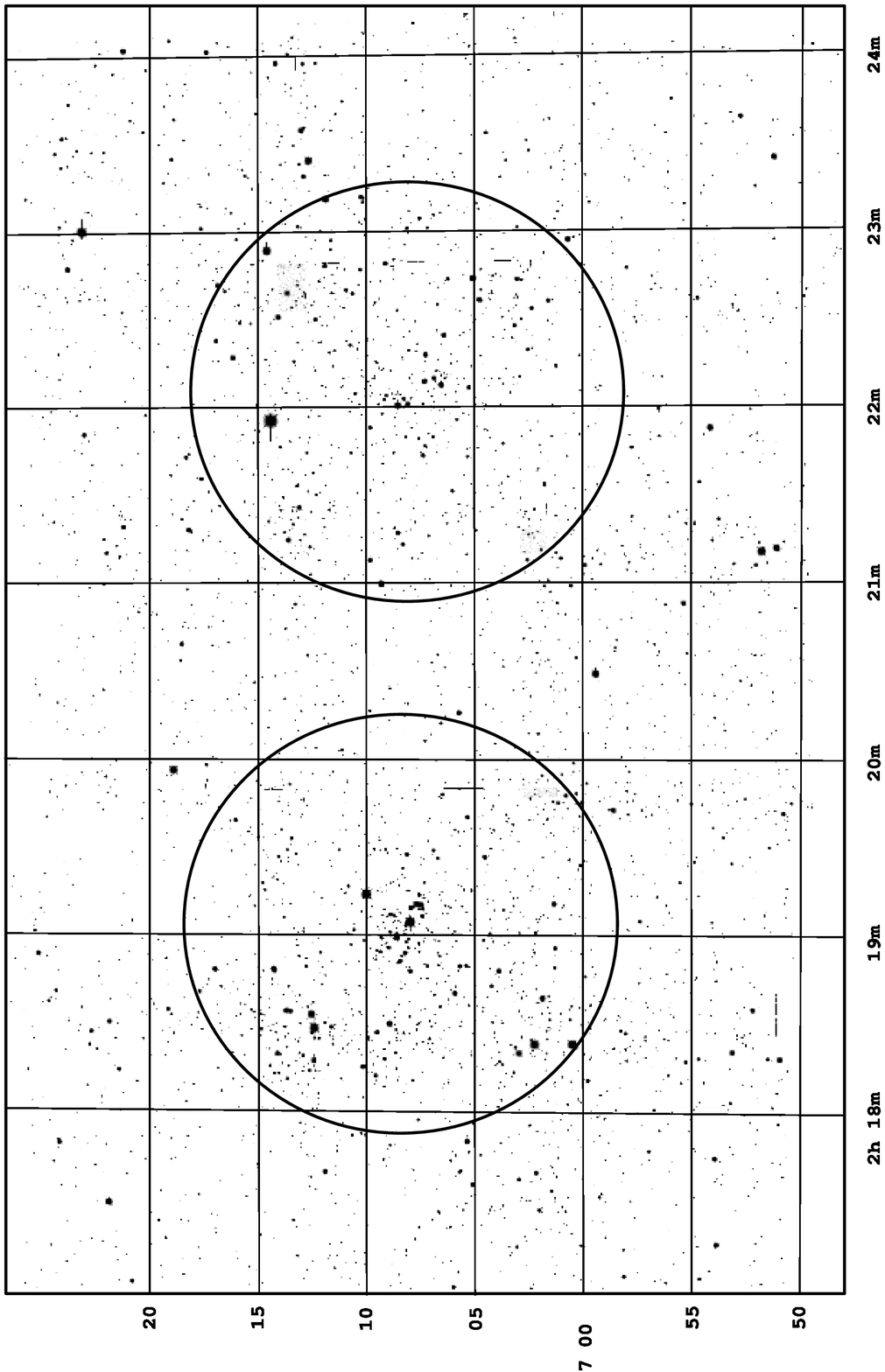}
\caption{Mosaic of the field of the present study. The circles mark
the boundaries taken for the extent of the individual clusters $h$ (left) \&
$\chi$ (right) Persei as discussed below.}
\label{image}
\end{figure}

$UBVI$ and H$\alpha$ photometry was performed on five nights between
17th and 25th of August 1999 at the San Diego State University Mount
Laguna Observatory with the 1m telescope and a Loral 2048 $\times$
2048 CCD. The scale was 0.$\arcsec$399 pixel$^{-1}$, yielding a field
of view of 13.$\arcmin$6$^{2}$. Fifteen frame centers were taken to
image the total field. The mosaiced 37\arcmin\ $\times$ 1\arcdeg\
field of our observations is shown in Figure \ref{image}.

The central wavelength and bandwidth of the H$\alpha$ filter were 6560
and 55\AA$\;$ respectively. Two sets of exposure times were used -
long: 360 s in H$\alpha$, 300 s $U$ and 60 s in $B V$ and $I$ - and
short: 40 s in $U$ and $I$, 20 s in $B$, 10 s in $V$. The mean
seeing for the observations was 1.\arcsec8. The observations were made
in photometric conditions.

Preprocessing, such as overscan correction, bias subtraction and
flat-fielding, was done using the IRAF CCDRED package. Instrumental
magnitudes were obtained using DAOPHOT II via point-spread function
fitting. The primary extinction coefficients and zeropoints were
determined every night from observations of Landolt standard regions
(Mark A, PG 1657+078, PG 2213-006, PG 1633+099: Landolt
1992). For the secondary extinction and transformation
coefficients we used the mean of five independent determinations
(errors represent the weighted mean errors). All the instrumental
magnitudes were transformed to the standard $UBVI$ system using the
following transformations:

\begin{equation}
V = v -k_{1,V}X + (0.006\pm0.007)(B-V) + \zeta_{V}
\end{equation}
\begin{equation}
B = b -[k_{1,B}X - (0.021\pm0.004)(B-V)]X-(0.053\pm0.004)(B-V)+\zeta_{B}
\end{equation}
\begin{equation}
U = u -[k_{1,U}X + (0.011\pm0.005)(U-B)]X-(0.120\pm0.013)(U-B) + \zeta_{U}
\end{equation}
\begin{equation}
I = i -k_{1,I}X + (0.063\pm0.012)(V-I) + \zeta_{I}
\end{equation}

where, following standard notation, $k_{1}$, $X$ and $\zeta$ represent
the primary extinction coefficient, air mass and zeropoint
respectively and the lowercase symbols denote the instrumental
magnitudes. Table \ref{tabcoef} lists the extinction coefficients and
zeropoints. In the case of the H$\alpha$ photometry only the
effects of atmospheric extinction were accounted for. Table
\ref{tabdata} presents our photometry.

The $UBV$ magnitudes and colours from this study are compared with the
previous photometry of Wildey \cite{wid64}, Moffat \& Vogt
\cite{mof74}, Johnson \& Morgan \cite{joh55}, Krzesinski et al.\
\cite{krz99}: NGC 864 nucleus) and Krzesinski et al.\ \cite{krz97}:
NGC 884 nucleus) in Table \ref{photcf}. The number of stars used in
the comparison is given in the fourth column by $n_{tot}$. Known
variables were excluded in the statistics. The photographic photometry
shows the largest scatter. The photometry of Wildey suffers greatly
from the effects of crowding, this results in the large dispersion and
generally brighter V magnitudes described in Table \ref{photcf}.

\section{Photometric Diagrams}

\begin{figure}
\plotone{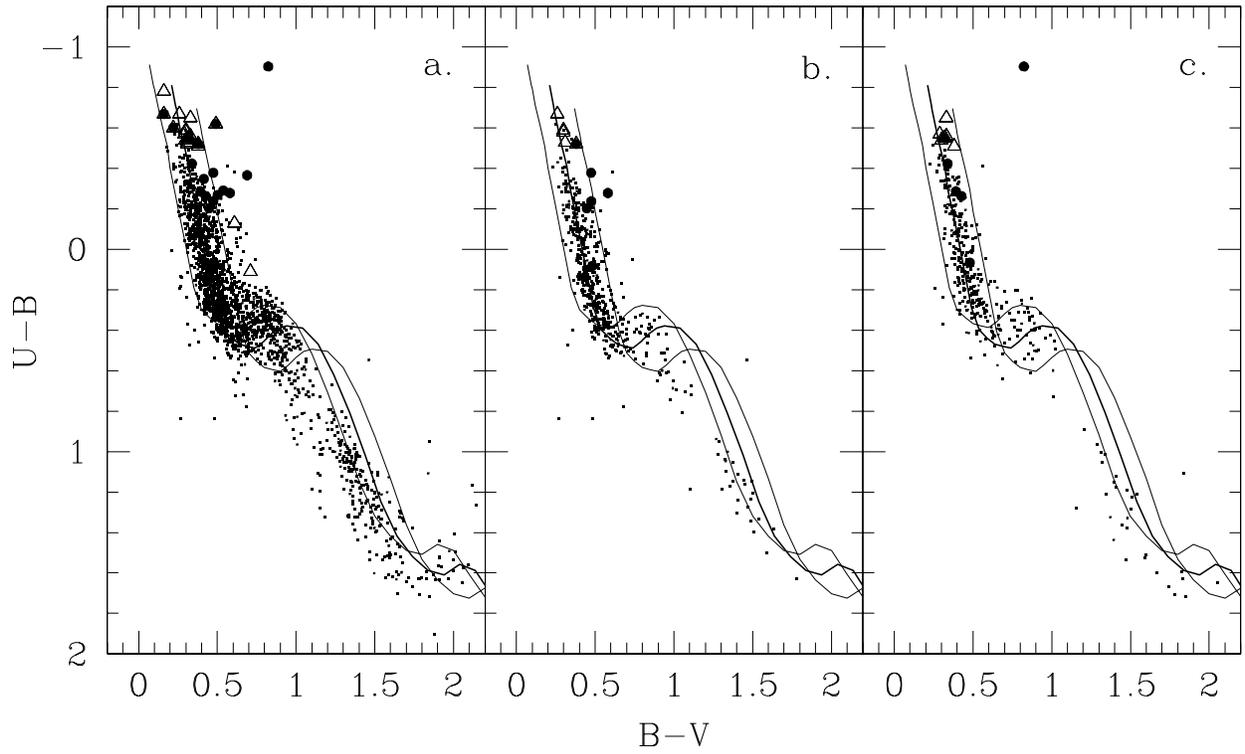}
\caption{$U$$-$$B$, $B$$-$$V$ diagram for (a) the entire field (b) $h$
Per. and (c) $\chi$ Per. The thick line represents the ZAMS relation
for $E(B$$-$$V)$ = 0.54, while the two thin lines are $E(B$$-$$V$) =
0.65 and 0.45 respectively. Filled circles are emission line stars and
triangles are stars with photometry taken from the literature.}
\label{umbbmvfig}
\end{figure}

In Figure \ref{umbbmvfig} we present the $U$$-$$B$, $B$$-$$V$
colour-colour (C-C) diagram from our photometry. In Fig.\
\ref{umbbmvfig} a.\ we show the entire field. In panels b and c we show
the C-C diagrams for $h$ and $\chi$ Per.\ contained within the
10\arcmin\ radius. The choice of this radius for both clusters
is made from a examination of the radial object density around each
cluster. At a radius of $\sim$10\arcmin\ the surface density
diminishes to that of the background field.

The reddened zero-age main sequence (ZAMS) is also shown in Fig.\
\ref{umbbmvfig} using the adopted distance moduli and mean reddenings
detailed in Sect. \ref{reddening}. The adopted ZAMS relation is from
Balona and Shobbrook \cite{bal84}. The dispersion seen about the mean MS
locus (solid line) in Fig.\ \ref{umbbmvfig} a.\ shows that there is
significant differential reddening across the field.

\begin{figure}
\plotone{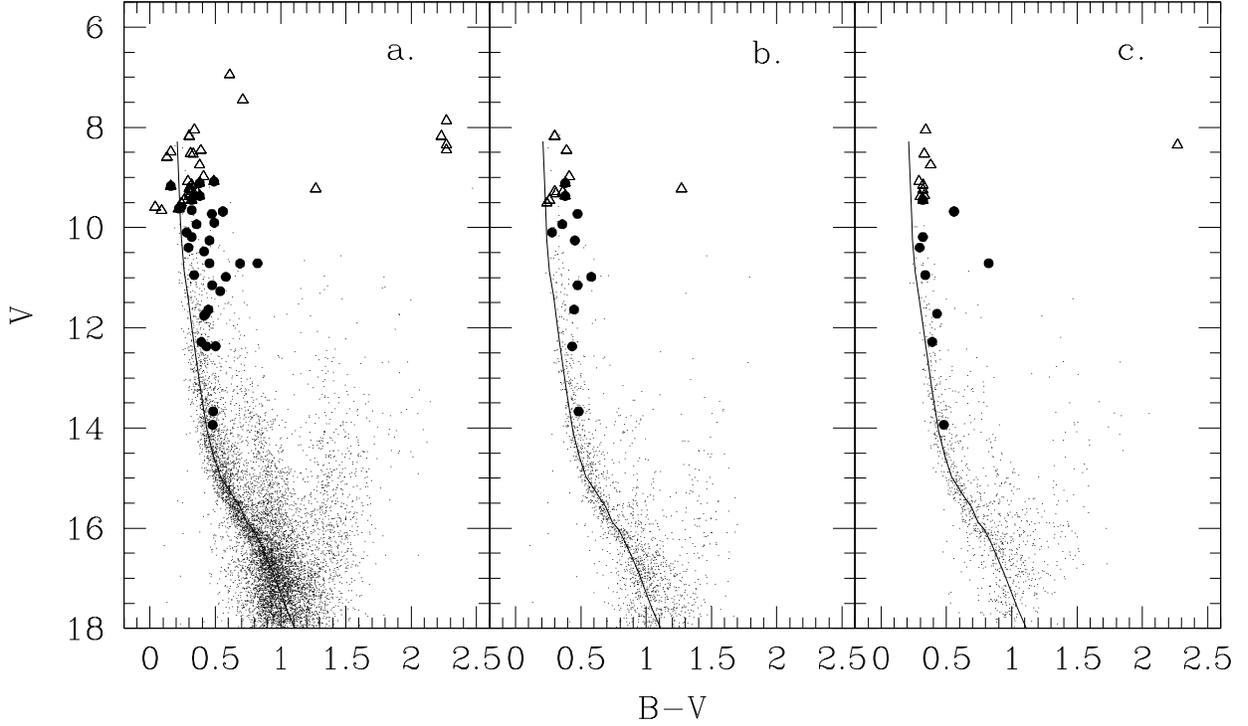}
\caption{$V$, $B$$-$$V$ diagram. Symbols are as described in Figure \ref{umbbmvfig}. The thin line is the ZAMS with $E$($B$$-$$V$) = 0.54.}
\label{vbmvfig}
\end{figure}

\begin{figure}
\plotone{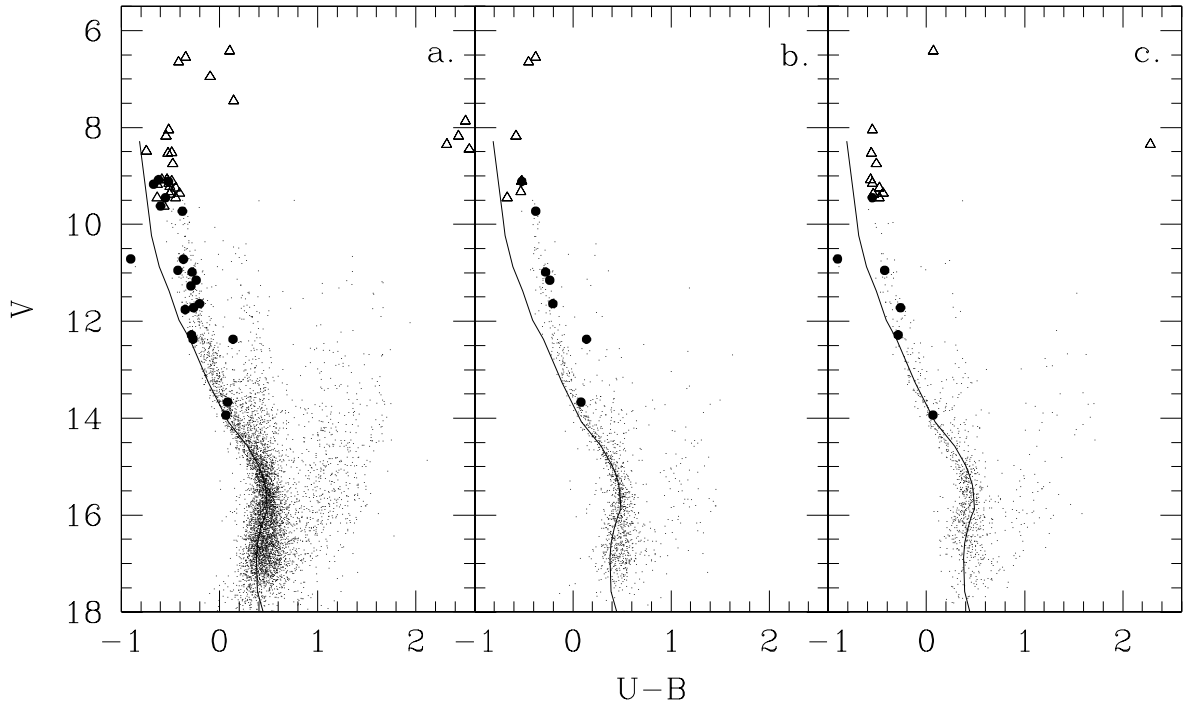}
\caption{$V$, $U$$-$$B$ diagram.}
\label{vumbfig}
\end{figure}

\begin{figure}
\plotone{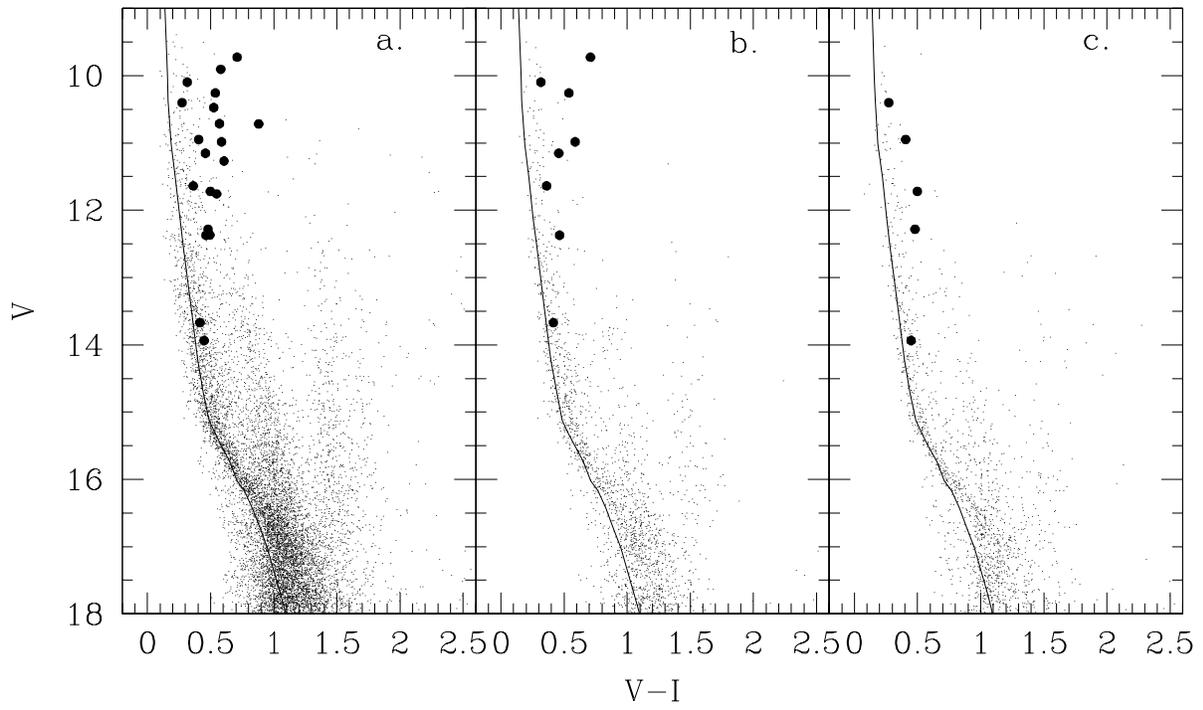}
\caption{$V$, $V$$-$$I$ diagram.}
\label{vvmifig}
\end{figure}

Figures \ref{vbmvfig}-\ref{vvmifig} show the $B$$-$$V$, $U$$-$$B$ and
$V$$-$$I$ colour-magnitude (C-M) diagrams for the entire field, $h$ and
$\chi$ Per.\ respectively. Our photometry is saturated in $B, V$ and
$I$ at 9.5. For stars brighter that $V\sim9.5$ we have relied on
previous photometry (Johnson \& Morgan 1955; Hiltner 1956; Wildey
1964; Tapia et al.\ 1984). Points for which we have utilized one or
more magnitudes from the literature are represented by triangular
symbols in Figs.\ \ref{umbbmvfig}-\ref{vvmifig}. Evolutionary
departure from the ZAMS is seen in Figs.\ \ref{vumbfig} and
\ref{vbmvfig}.

\section{Reddening and Distance Moduli}
\label{reddening}

We adopt the nominal values for colour-excess ratio
$E(U-B)/E(B-V)=0.72$ and the ratio of total to selective extinction,
$R$ ($\equiv A_{V}/E(B-V)$), of 3.2 (see e.g.\ Schmidt-Kaler
1982). Using those early MS, non-emission line stars
($U$$-$$B<0.15$ and $B-V<0.8$ in Figure \ref{umbbmvfig}) we derived
$E(B-V)$ using the Q technique ( Q$ = (U-B) - 0.72(B-V)$; see
e.g. Bessell 1991 ): E($B-V)=(B-V)_{obs} - Q/3.2$, where
($B$$-$$V$)$_{obs}$ is the observed colour.

\begin{figure}
\plotone{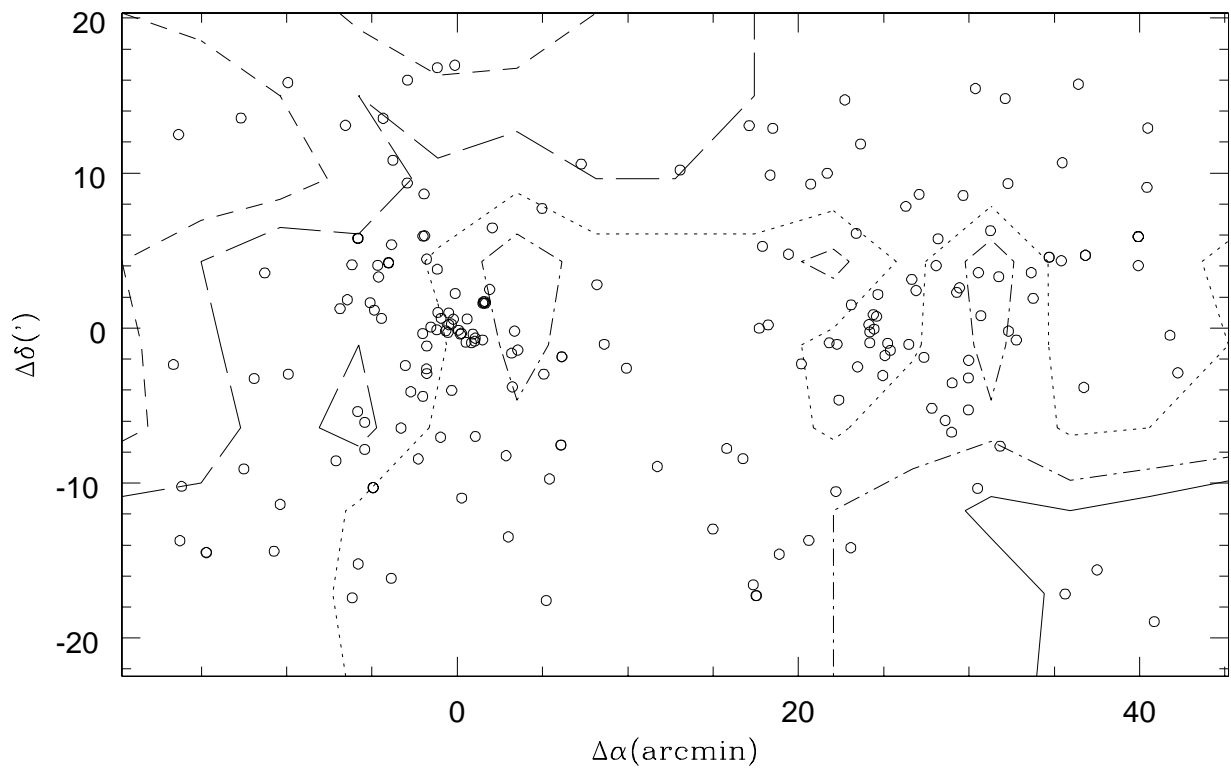}
\caption{Reddening distribution in the vicinity of $h$ \& $\chi$
Persei. Positions relative to the center of $h$ Persei at 02:19:00
+57:09:00 (J2000). The 100 brightest objects are superimposed. The
solid, dot-dashed, dotted, long-dashed and dashed lines are
respectively $E(B-V)$= 0.65, 0.60, 0.55, 0.50 and 0.45.}
\label{ebmvfig}
\end{figure}

In order to describe the spatially variable reddening our sample was
divided into a grid of regions such that each region contained at
least 10 stars for which we have determined $E(B-V)$. The resulting
spatial distribution of $E(B-V)$ is shown in Figure \ref{ebmvfig}. The
region possesses a clear negative reddening gradient from SW to NE which is
consistent with increasing galactic latitude. The nuclei of both
clusters have average reddenings of $E(B-V)=0.54\pm0.02$ from 96 and
85 stars within $h$ \& $\chi$ Per. respectively. This is in agreement
with the determination of Balona \& Shobbrook \cite{bal84} from the
$UBV$ photometry of Crawford et al.\ \cite{cra70} of a mean
reddening of $E(B-V)=0.559\pm0.009$. Each star is then de-reddened on
the basis of its position within Figure \ref{ebmvfig}.

Once de-reddened, the ZAMS of Balona and Shobbrook \cite{bal84} was
used to determine the distance modulus of each star. The bulk of the
MS A \& F stars ($V_{0} \sim 12.5$-15) are well matched by the ZAMS.
Significant evolutionary departures occur for the early B stars, these
were excluded from our calculation. The distribution of distance
moduli within the two clusters shows a broad peak centered at
$V_{0}-M_{V}=11.75\pm0.05$. There is no significant difference in
distance modulus of the two clusters. This result uses the bulk
of the essentially unevolved MS. It is significantly more distant than
the distance moduli derived by Balona \& Shobbrook \cite{bal84} of
$V_{0}-M_{V}$=$11.16\pm0.08$ which is derived from a limited sample of
upper MS stars rectified to a similar sample of stars in the
Pleiades. It is within the range of uncertainties of the Marco \&
Bernabeu \cite{mar01} result of dm=11.66$\pm$0.20 and 11.56$\pm$0.20
for $h$ \& $\chi$ Persei respectively.

\section{The Be Star Population}
\label{secbe}

\begin{figure}
\plotone{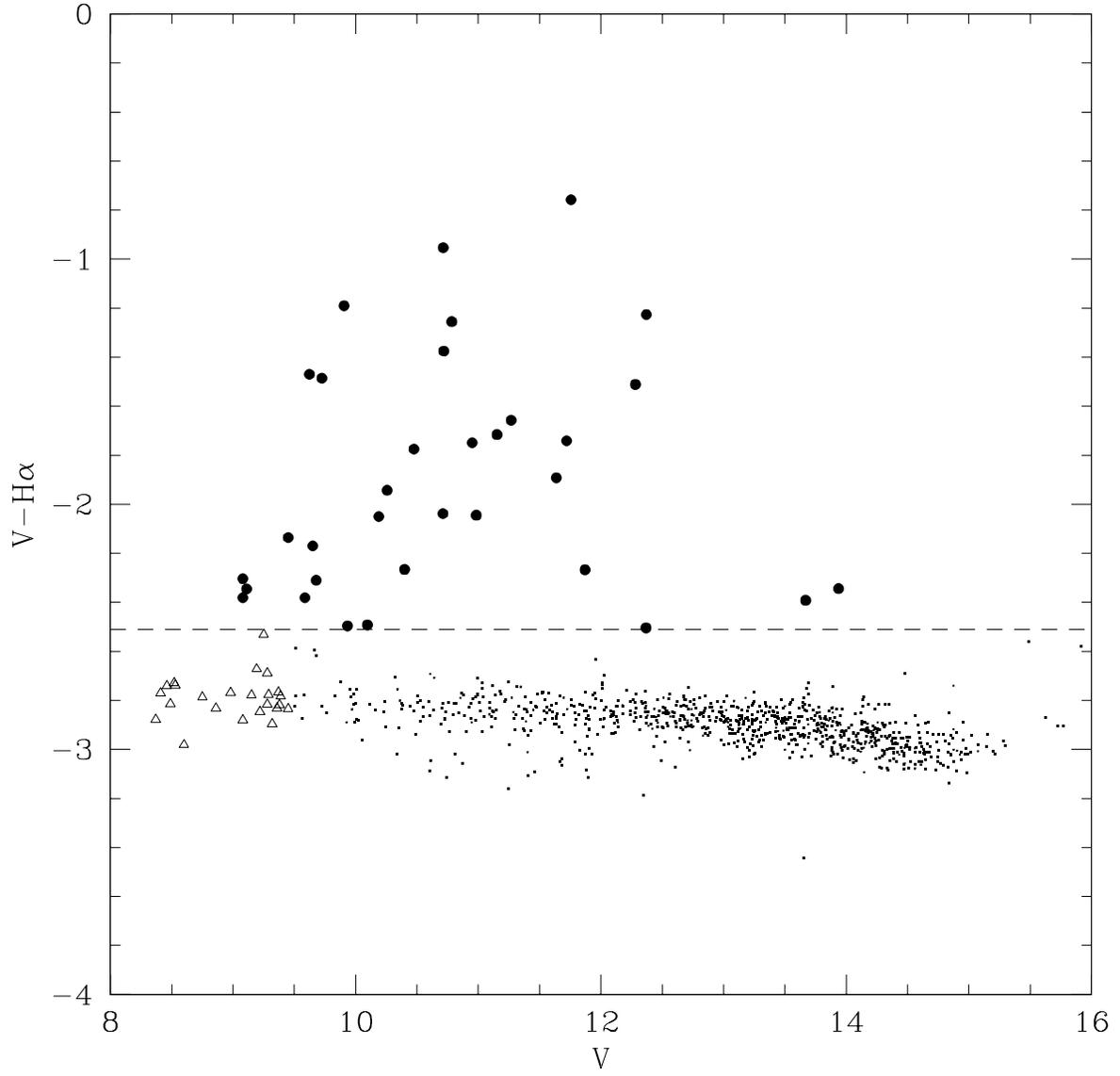}
\caption{$V$, $V$$-$H$\alpha$ C-M diagram. Those objects above the
dashed line show significant H$\alpha$ emission.}
\label{vhafig}
\end{figure}

For the purposes of selecting the population of emission-line objects
we construct a $V$,$V$$-$H$\alpha$ C-M diagram (Fig. \ref{vhafig}). We
construct such a diagram in preference to the $V$$-$$I$,
$V$$-$H$\alpha$ C-C diagram of previous studies (see Grebel
et al.\ 1992 and Grebel 1997). We find a total of 33
emission-line objects within the surveyed field as described in Table
\ref{tabbe}. The total from previous studies within this area is
25. Two known Be stars, Oo 566 and Oo 2262 were not found to exhibit
significant emission on the epoch of our observations. We do not
detect any faint emission-line objects which can be clearly associated
with a pre-MS population.

\section{Hertzsprung-Russell Diagram}
\label{secage}

\subsection{Transformation to the H-R Diagram}

In order to form the H-R diagram for the cluster populations we have
made use of the model atmosphere broad-band colours, bolometric
corrections and temperatures of Bessell, Castelli \& Plez
\cite{bes98}. For stars of high effective temperature the $UBV$
colours fail to provide good temperature indicators (Massey et al.\
1995) and consequently we have used spectral types for
temperature determination where possible for stars earlier than
B2. These spectral types are drawn from Johnson \& Morgan
\cite{joh55}, Schild (1965; 1967) and Slettebak
\cite{sle68}. In doing so we have used the spectral type - effective
temperature conversion of B\"ohm-Vitense \cite{boh81}. For
later-type stars ($U$$-$$B$)-T$_{eff}$ and ($B$$-$$V$)-T$_{eff}$
relations were used with weights reflective of the temperature
dependence to the colour of the star in question. Using the above
relations the H-R diagram for $h$ \& $\chi$ Persei is constructed as
shown in Figure \ref{fighrd}.

\subsection{Cluster Ages}
\begin{figure}
\plotone{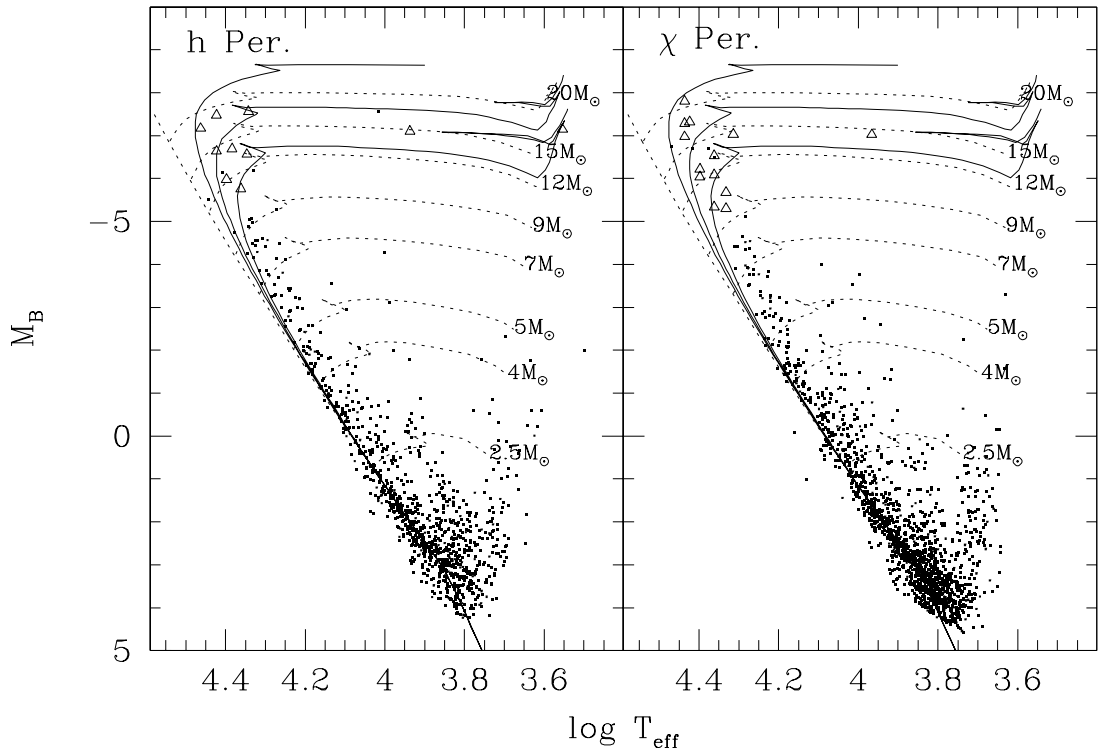}
\caption{The H-R diagram of $h$ and $\chi$ Persei. The empty triangles
are points derived from spectral types. The population of Be stars is
excluded from this figure together with those stars with membership
probabilities of less than 20\% (Mumimov 1983). Isochrones (solid lines) are shown for log age(yr) = 6.8, 7.0 \& 7.2.}
\label{fighrd}
\end{figure}

In Figure \ref{fighrd} we have removed those stars which are highly
likely to be non-members according to the proper motion study of
Mumimov \cite{mum83}: with probability of membership less than
20\%). Be stars were also excluded due to their anomalous colours
produced by the presence of circumstellar material.  This
clarification is particularly useful in the vicinity of the MS
turnoff. In Figure \ref{fighrd} we overlay stellar models by Bressan
et al.\ \cite{bre93} and isochrones derived from them (Bertelli et
al.\ 1994). The ages of the most massive stars lie between the
isochrones of log age(yr)=6.8 and 7.2. Together with the evolved members
the populations suggest a common age of log age(yr)=7.1$\pm$0.1 for the
bulk of the population in both clusters, in line with the studies of
Waelkens et al.\ \cite{wae90} and Denoyelle et al.\
\cite{den90}. This is in disagreement with the findings of Marco \&
Bernabeu \cite{mar01} who, from an examination of the average MS
locus of both clusters, find that $h$ Persei with a log age of 7.3 is
older than $\chi$ Persei (with a log age of 7.0). Marco \& Bernabeu go
further to propose three epochs of star formation within $h$ \& $\chi$
Persei at log ages 7.0, 7.15 and 7.3. We propose that the analysis of
Marco \& Bernabeu \cite{mar01} is largely an over-interpretation of
the available data. Our expanded dataset does not show any evidence
for an age difference between the two clusters or multiple epochs of
star formation.

It would be useful to consider the mass functions of the two clusters
together with the age structure of the surrounding field. At present
this is not possible. The existing proper motion study of Mumimov
\cite{mum83} is limited in magnitude ($V \leq 12$) and spatial extent
(only the cluster cores are covered). There is no photometric method
for distinguishing the cluster members from members of the surrounding
association due to the close affinity in age of the cluster and field
populations. Such analysis awaits more detailed proper motion studies.

\section{The Be Star Fraction - Evidence for an Evolutionary Enhancement in Be Phenomenon?}

\begin{figure}
\plotone{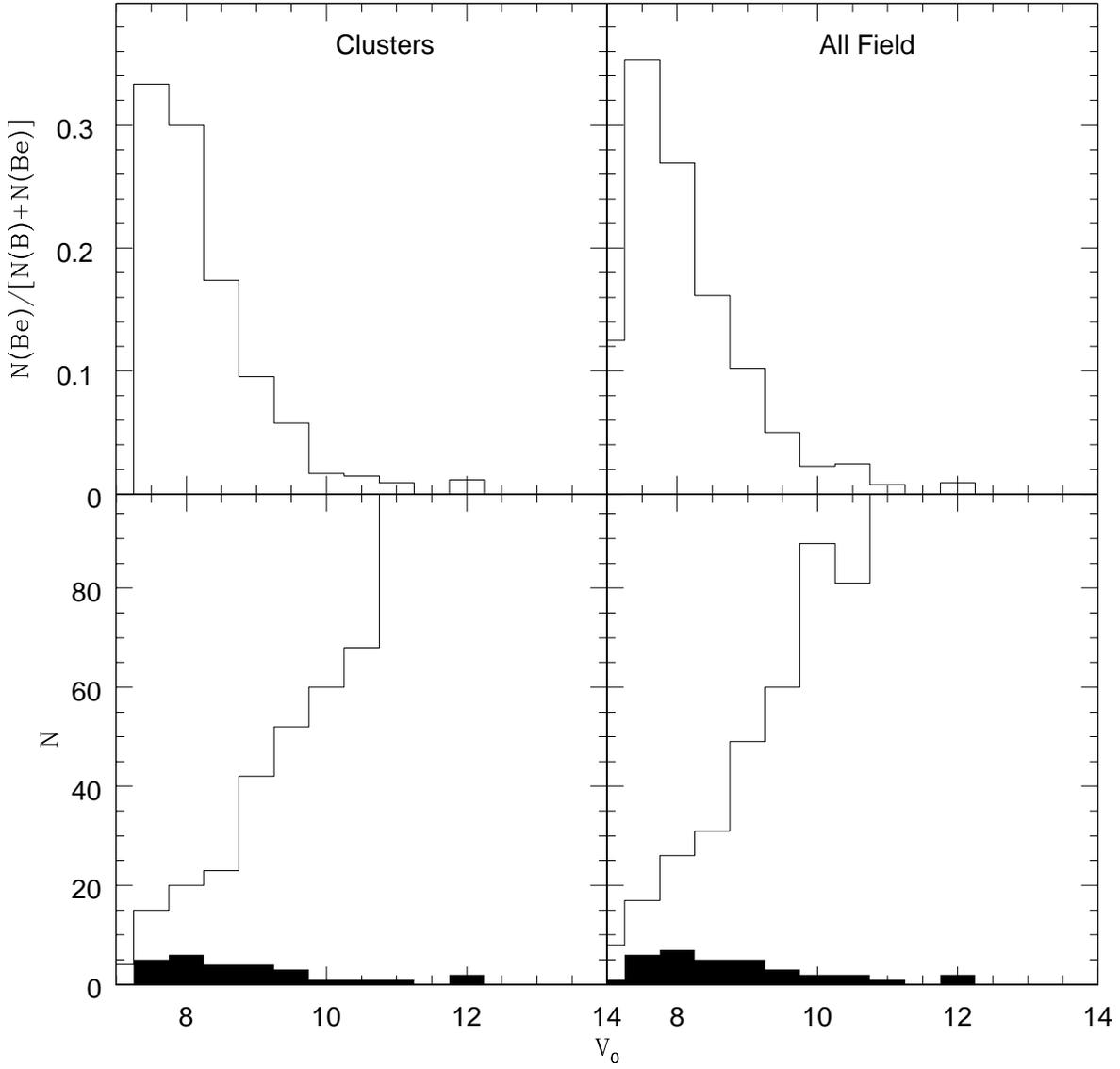}
\caption{Top panel shows the fraction of Be star to total stars within
the cluster population and the entire field. Bottom panel shows the
number of B and Be (shaded) stars within both regions.}
\label{beffig}
\end{figure}

In Figure \ref{beffig} we present the Be fraction amongst MS B stars
(to $M_{V}$=+0.25 $\equiv$ B9V Schmidt-Kaler 1982) within the
field of the association. The resulting distribution peaks at a
fraction of 36\% over the 0.5 mag. interval centered on $V_{0}$=7.5
which is in the vicinity of the MS turnoff for the association.

The peak in Be fraction seen at the MS terminus is also seen in
several young clusters of the Magellanic Clouds (Keller et al.\
2000 and Johnson et al.\ 2001). Yet studies of the Be
fraction within the Galactic and LMC field (see Zorec \& Briot
1997 and Keller et al.\ 1999b respectively) show a
more or less uniform proportion of around 15\% over the spectral range
B0-B5. To account for the observed difference in Be fraction between
the cluster and field environment we propose a scenario of
evolutionary enhancement in the Be phenomenon. Our scenario utilises
the one clear difference between the cluster and field populations,
namely the age spread within each population.

We consider a phase of enhancement of the Be phenomenon which occurs
over an interval of the MS lifetime towards the end of the MS. The
small age spread within the cluster population at the luminosity of
the cluster MS terminus places the majority of the stars within the
interval of Be enhancement. A peak in the Be fraction results at
luminosities near the MS terminus and diminishes at lower luminosities
as fewer stars reside within the age range of enhancement. At lower
luminosities a population of Be stars remains. These stars are those
with sufficient initial angular momentum to commence the Be
phase. Within the field population, which possesses a more uniform
spread of ages for stars on the MS at a given mass, we see essentially
the time-average of the proportion of Be stars. This is lower than
that seen in the cluster and more-or-less uniform with luminosity.

The tendency for there to be a higher Be star fraction near the
main-sequence turnoff is consistent with the rapid rotation hypothesis
for Be star formation.  The evolution of rotating stars has been
examined by Endal \& Sofia \cite{end79} and recently by Heger et
al.\ \cite{heg00} and Meynet \& Maeder \cite{mey00}. These models
explicitly follow the radial exchange of angular momentum with
evolution. The quantitative details of the evolution of the equatorial
surface velocity ($v$) depend critically on mass loss. Taking the
example from Meynet \& Maeder \cite{mey00}, a 20M$_{\odot}$ star
with an initial $v$ of 300 kms$^{-1}$ drops to $\sim$120 kms$^{-1}$ by
the MS turnoff. On the other hand a 12M$_{\odot}$ star drops from 300
kms$^{-1}$ to $\sim$200 kms$^{-1}$ due to a substantially smaller mass
lose rate. The evolution of $\frac{\Omega}{\Omega_{crit}}$, the
surface angular velocity as a fraction of the critical angular
velocity (i.e.\ the velocity at which the equatorial escape velocity drops
to zero), is dependent on the treatment of redistribution of angular
momentum within the stellar interior and of the effects of
rotationally induced mixing. Heger et al.\ and Meynet \& Maeder
implement different treatments of the above but the qualitative
conclusions are concordant: for stars less than 15M$_{\odot}$,
$\frac{\Omega}{\Omega_{crit}}$ increases towards the end of the MS
lifetime.

Meynet \& Maeder examine the case of a 20M$_{\odot}$ star with a range
of initial velocities. They find that the decrease in the surface $v$
is larger for larger initial velocities. This is due to the high
stellar mass loss rates for rapid rotators. At lower masses we expect
this behavior to be modified by lower mass loss rates. The models of
Endal and Sofia examine the rotational velocity evolution of a
5M$_{\odot}$ star. They find that stars commencing their main-sequence
lives with relatively slow angular velocity remain slow rotators
throughout the whole of the main-sequence evolutionary phase (although
there is a marked increase in angular velocity of all stars during the
core contraction phase associated with exhaustion of hydrogen in the
core, this evolutionary phase is far too short to account for the
observed proportion of Be stars as was proposed by Schmidt-Kaler
\cite{sk65} and recently by Bessell \& Wood \cite{bes93}.  Thus slow
rotators will never evolve into Be stars.

In contrast to the slow rotators, the models of Endal and Sofia show
that stars commencing their lives with an angular velocity greater
than 56-76\% of the critical breakup velocity $\Omega_{crit}$ spin up
to the critical velocity over a moderate fraction of the main-sequence
lifetime.  In a cluster where some fraction of the stars is formed
with angular velocity greater than $\sim$50\% of $\Omega_{crit}$,
those stars nearer the main-sequence turnoff are more likely to be Be
stars since they will have been more spun up by evolution than the
less luminous stars that have not evolved as far through the
main-sequence phase. This process could provide the mechanism of Be
enhancement required to explain the concentration of Be stars to the
main-sequence tip.

Numerous studies have indicated the importance of rotation within
massive stars (e.g. Langer \& Heger 2000 and Meynet \& Maeder
2000). Rotationally induced mixing has been shown to be able to
produce significant evolutionary and surface abundance
changes. Observational evidence for the importance of rotation is
growing, within $h$ \& $\chi$ Persei in particular, the study of
Vrancken et al.\ \cite{vra00} provides a detailed chemical analysis
of eight B1 - B2 stars within the vicinity of the MS turnoff. They
find, from the deduced HR diagram, a degree of extension to the end of
the MS which is in agreement with evolutionary models which
incorporate rotation.

\section{Summary} 

We have obtained $UBVI$ and H$\alpha$ photometry over a 37\arcmin\
$\times$ 1\arcdeg\ field around the double cluster $h$ \& $\chi$
Persei. We have formed a spatial description of the reddening
throughout the field. We have found that the clusters share a common
distance modulus of 11.75$\pm$0.05 and possess ages of log
age(yr)=7.1$\pm$0.1. Our study comprises a uniform survey for Be stars
within the region. Using the $V$$-$H$\alpha$ colour as a measure of
the strength of H$\alpha$ emission we have identified 33 Be stars, 8
of which were previously unknown. The fraction of Be stars to total B
stars shows a maximum at the end of the MS. Similar behaviour is seen
in the Be population of young clusters in the Magellanic Clouds. This
feature suggests a evolutionary enhancement in the Be phenomenon
towards the end of the MS lifetime which we suggest is a consequence
of rotational spin-up over the MS lifetime.

\begin{acknowledgements} 
This work was performed under the auspices of the U.S. Department of
Energy by the University of California Lawrence Livermore National
Laboratory under contract No. W-7405-Eng-48. The authors acknowledge
the use of the WEBDA database maintained by J-C. Mermilliod,
Univ. Lausanne. SCK acknowledges funding from the Australian Nuclear
Science and Technology Organisation which made the observations
possible.
\end{acknowledgements}

\clearpage

\begin{table} 
\begin{center} 
\begin{tabular}{lcccc} 
\hline 
\hline 
Date& $k_{1,V}$ & $k_{1,B}$ & $k_{1,U}$ & $k_{1,I}$ \\ 
\hline 
Aug 17 & $0.147\pm0.015$ & $0.142\pm0.021$ & $0.27\pm0.06$ & $0.034\pm0.022$\\
Aug 18 & $0.135\pm0.011$ & $0.158\pm0.042$ & $0.32\pm0.07$ & $0.04\pm0.03$\\
Aug 19 & $0.150\pm0.017$ & $0.145\pm0.025$ & $0.43\pm0.04$ & $0.07\pm0.03$\\
Aug 21 & $0.138\pm0.025$ & $0.152\pm0.031$ & $0.29\pm0.09$ & $0.04\pm0.05$\\
Aug 23 & $0.159\pm0.031$ & $0.163\pm0.022$ & $0.42\pm0.04$ & $0.08\pm0.03$\\
\hline 
\hline 
Date& $\zeta_{V}$ & $\zeta_{B}$ & $\zeta_{U}$ & $\zeta_{I}$ \\ 
\hline 
Aug 17 & $2.235\pm0.025$ & $2.522\pm0.018$ & $0.05\pm0.08$ & $1.63\pm0.04$\\
Aug 18 & $2.493\pm0.03$ & $2.690\pm0.023$ & $0.11\pm0.08$ & $1.35\pm0.03$\\
Aug 19 & $2.482\pm0.03$ & $2.36\pm0.03$ & $0.09\pm0.06$ & $1.60\pm0.04$\\
Aug 21 & $2.883\pm0.04$ & $2.794\pm0.05$ & $0.03\pm0.09$ & $1.74\pm0.06$\\
Aug 23 & $2.539\pm0.03$ & $2.473\pm0.024$ & $0.13\pm0.05$ & $1.83\pm0.05$\\
\hline 
\end{tabular} 
\end{center} 
\caption{Atmospheric extinction coefficients and photometric zeropoints from the present study} 
\label{tabcoef} 
\end{table} 

\begin{table} 
\begin{center} 
\begin{tabular}{lcccccccc} 
\hline 
\hline 
ID  & $x$ & $y$ & $V$ & $U-B$ & $B-V$ & $V-I$ & $V-H\alpha$& Oost. ID\\ 
\hline 
   100& 2184.388&  944.050& 10.380& -0.314&  0.464&  0.412& -2.822&2794\\
   101& 2203.848& -229.542& 10.475&  \ldots& 0.414&  0.526& -1.605&2804\\
   102&  171.524& -493.646& 10.400&  0.236&  0.523&  0.453& -2.786&1242\\
   103&  324.182& -584.171& 10.443& -0.411&  0.261&  0.144& -2.823&1391\\
   104&  -56.437&   38.175& 10.477& -0.387&  0.310&  0.232& -2.845& 936\\
   105& 1109.442&  772.894& 10.470&  0.110&  0.567&  0.443& -2.774&1932\\
\hline 
\end{tabular} 
\end{center} 
\caption{Photometric data. Table \ref{tabdata} is presented in its entirety in the electronic edition of the Astronomical Journal. A portion is shown here for guidance regarding its form and content.} 
\label{tabdata} 
\end{table} 

\begin{table} 
\begin{center} 
\begin{tabular}{lccccc} 
\hline 
\hline 
Author & Method & $\Delta V$ & $n_{tot}$ & $\Delta(B-V)$ & $\Delta(U-B)$\\ 
\hline 
Johnson \& Morgan \cite{joh55}& photoelectric & $-0.009\pm0.068$ & 76 & $0.000\pm0.053$ & $0.005\pm0.076$\\ 
Wildey \cite{wid64}& photographic & $-0.16\pm0.38$ & 243 & $0.03\pm0.11$ & $0.00\pm0.18$\\ 
Moffat \& Vogt \cite{mof74} & photographic & $-0.01\pm0.08$ & 1437 & $-0.02\pm0.14$ & $0.03\pm0.17$\\ 
Krzesinski et al.\ \cite{krz97}& CCD & $-0.023\pm0.061$ & 202 & $-0.041\pm0.087$ & $-0.050\pm0.057$\\ 
Krzesinski et al.\ \cite{krz99}& CCD & $0.001\pm0.063$ & 68 & $-0.017\pm0.059$ & $0.025\pm0.12$\\ 
\hline 
\end{tabular} 
\end{center} 
\caption{Comparison with previous photometric studies. $\Delta$ is in
the sense of previous study minus the value of obtained in the present
study.}
\label{photcf} 
\end{table} 

\begin{table} 
\begin{center} 
\begin{tabular}{lcccccccc} 
\hline 
\hline 
ID  & $x$ & $y$ & $V$ & $U-B$ & $B-V$ & $V-I$ & $V-H\alpha$& Oost. ID\\ 
\hline 
    28& 1363.207&  883.212&  9.080& -0.620&  0.490&  \ldots& -2.304&2138\\
    29& -120.888& -264.748&  9.110& -0.520&  0.380&  \ldots& -2.345& 847\\
    31& -644.392& -864.324&  9.170& -0.670&  0.160&  \ldots& -1.923& 309\\
    42&  201.864&  -11.658&  9.370& \ldots&  0.380&  \ldots& -2.466&1268\\
    45& 1308.504&  -56.666&  9.450& -0.550&  0.320&  \ldots& -2.136&2088\\
    54& 1301.495&  599.813&  9.587& \ldots&  0.039& \ldots & -2.381&2079\\
    55& -881.546& -869.430&  9.620& -0.600&  0.220& \ldots  & -2.419& 146\\
    56&  784.349&  611.905&  9.650& \ldots&  0.320& \ldots & -2.170&1702\\
    58& 1624.853&  517.354&  9.678& \ldots&  0.558& \ldots & -2.310&2402\\
    60& 1496.850& -183.588&  9.694& -0.512&  0.425& \ldots & -1.744&2284\\
    62&  193.788& -227.629&  9.725& -0.378&  0.474&  0.709& -1.486&1261\\
    66& 1100.835&  591.965&  9.905& \ldots&  0.493&  0.580& -1.190&1926\\
    68& -114.916&  357.033&  9.934& \ldots &  0.357& \ldots & -2.496& 846\\
    78& -386.925&  110.086& 10.096& \ldots&  0.280&  0.316& -2.492& 517\\
    82& 1385.516&   89.806& 10.189& -0.563 &  0.319& \ldots & -2.050&2165\\
    88&   89.547&  -45.684& 10.257& \ldots&  0.454&  0.538& -1.943&1161\\
    91& 1966.636&  -45.877& 10.399& \ldots&  0.295&  0.273& -2.266&2649\\
   101& 2203.848& -229.542& 10.475& \ldots&  0.414&  0.526& -1.775&2804\\
   126& 1833.016&  214.942& 10.712& -0.621&  0.623& \ldots& -2.038&2566\\
   132& 1830.679& -620.322& 10.714& \ldots &  0.454&  0.571& -0.953&2563\\
   137& 2128.076&  640.403& 10.718& -0.366&  0.689&  0.881& -1.205&2759\\
   149& 1463.242&   51.723& 10.950& -0.424&  0.338&  0.406& -1.749&2242\\
   163&  212.904&  -85.200& 10.984& -0.278&  0.580&  0.587& -2.045&1282\\
   191&  367.814& -110.785& 11.152& -0.237&  0.475&  0.458& -1.716& \ldots\\
   209& 2151.376&  902.980& 11.268& -0.291&  0.538&  0.606& -1.657&2771\\
   273&  214.032&  174.599& 11.637& -0.203&  0.447&  0.362& -1.892&1278\\
   279& 1316.548&  142.873& 11.720& -0.264&  0.427&  0.498& -1.741&2091\\
   285& -502.380&-1023.390& 11.756& -0.349&  0.415&  0.548& -0.758& 420\\
   381& 1157.409&  202.441& 12.281& -0.286&  0.392&  0.480& -1.551&1977\\
   409&   47.680& -280.088& 12.370&  0.138&  0.433&  0.464& -1.226&1114\\
   416& 1904.465&  913.611& 12.368& -0.268&  0.502&  0.494& -2.504&2600\\
   964&   18.442&   11.566& 13.669&  0.083&  0.483&  0.416& -2.392&1058\\
  1199& 1081.700&  423.908& 13.938&  0.066&  0.480&  0.449& -2.344&1912\\
\hline 
\end{tabular} 
\end{center} 
\caption{Be stars detected within the field of $h$ \& $\chi$ Persei.} 
\label{tabbe} 
\end{table}


\clearpage 

\begin{thebibliography}{}
\bibitem[1984]{bal84} Balona, L.\ A. \& Shobbrook, R.\ R. 1984, MNRAS, 211, 375
\bibitem[1994]{ber94} Bertelli G., Bressan A., Chiosi C., Fagotto F., Nasi E., 1994, A\&AS, 106, 275
\bibitem[1991]{bes91} Bessell, M.\ S., 1991, A\&A, 242, L17
\bibitem[1993]{bes93} Bessell, M.\ S.\ \& Wood, P.\ R.\ 1993, in New
Aspects of Magellanic Cloud Research, 271
\bibitem[1998]{bes98} Bessell, M.\ S., Castelli, F. \& Plez, B.\ 1998, A\&A, 333, 231
\bibitem[1981]{boh81} B\"ohm-Vitense, E. 1981, ARA\&A, 18, 241
\bibitem[1993]{bre93} Bressan A., Fagotto F., Bertelli G., Chiosi, C. 1993, A\&AS, 100, 647
\bibitem[1970]{cra70} Crawford, D.\ L., Glaspey, J.\ W. \& Perry, C.\ L. 1970, AJ, 75, 822
\bibitem[1990]{den90} Denoyelle, J.\ et al.\ 1990, Ap\&SS, 169, 109 
\bibitem[1979]{end79} Endal, A.\ S. \& Sofia, S. 1979, ApJ, 232, 531
\bibitem[1996]{fab96} Fabregat, J., Torrejon, J.\ M., Reig, P., Bernabeu, G., Busquets, J., Marco, A., \& Reglero, V.\ 1996, A\&AS, 119, 271 
\bibitem[1992]{gre92} Grebel, E.\ K., Richter, T., \& de Boer, K.\ S.\ 1992, A\&A, 254, L5 
\bibitem[1997]{gre97} Grebel, E.\ K.\ 1997, A\&A, 317, 448 
\bibitem[2000]{heg00} Heger, A., Langer, N. \& Woosley, S.\ E. 2000, ApJ, 528, 368
\bibitem[1956]{hil56} Hiltner, W.\ A.\ 1956, ApJS, 2, 389 
\bibitem[1955]{joh55} Johnson H.\ L., Morgan W.\ W. 1955, ApJ, 122, 429
\bibitem[2001]{joh01} Johnson R.\ A., Beaulieu, S.\ F., Gilmore, G.\ F., Hurley J., Santiago B.\ X., Tanvir N.\ R., Elson, R.\ A.\ W. 2001, astro-ph/0012389
\bibitem[1999a]{kel99a} Keller, S.\ C., Wood, P.\ R. \& Bessell, M.\ S. 1999, A\&AS, 134, 489
\bibitem[1999b]{kel99b} Keller, S.\ C., Bessell, M.\ S., \& Da Costa, G.\ S. 1999 in IAU Coll.\ 175, ed. M.\ Smith, H.\ Henrichs \& J.\ Fabregat (San Francisco: ASP), 141
\bibitem[2000]{kel00} Keller, S.\ C., Bessell, M.\ S., \& Da Costa, G.\ S. 2000, AJ, 119, 1748
\bibitem[1997]{krz97} Krzesinski, J. \& Pigulski, A. 1997, A\&A, 325, 987
\bibitem[1999]{krz99} Krzesinski, J. \& Pigulski, A. \& Kolaczkowski, Z. 1999, A\&A, 345, 505
\bibitem[1992]{lan92} Landolt, A.\ U.\ 1992, AJ, 104, 340 
\bibitem[1999]{mae99} Maeder, A., Grebel, E.\ K. \& Mermilliod, J-C.\ 1999, A\&A, 346, 459
\bibitem[2001]{mar01} Marco, A. \& Bernabeu, G. 2001, A\&A, accepted astro-ph/0102267
\bibitem[1995]{mas95} Massey, P., Lang, C., DeGioia-Eastwood, K. \& Garmany, C.\ 1995, ApJ, 438, 188
\bibitem[2000]{mey00} Meynet, G. \& Maeder, A. 2000, A\&A, 361, 101
\bibitem[1974]{mof74} Moffat, A.\ F.\ J. \& Vogt, N. 1974, Veroeff. Astron. Inst. Bochum 2
\bibitem[1953]{mor53} Morgan W.\ W. 1953, ApJ, 118, 318 
\bibitem[1983]{mum83} Muminov M. 1983 Bull. Inform. CDS, 24, 95 
\bibitem[1937]{oos37} Oosterhoff, P.\ T. 1937, Ann. van de Sterrewacht to Leiden, 17, 1
\bibitem[1965]{sch65} Schild, R.\ E. 1965, ApJ, 141, 979
\bibitem[1967]{sch67} Schild, R.\ E. 1967, ApJ, 148, 449
\bibitem[1965]{sk65} Schmidt-Kaler, T.\ 1965, AJ, 70, 691 
\bibitem[1982]{sch82} Schmidt-Kaler, T. 1982, in Landolt-B\"ornstein, Vol 2, 23, eds. L.H. Aller et al.\ (2d ed.; Berlin: Springer-Verlag)
\bibitem[1968]{sle68} Slettebak, A.\ 1968, ApJ, 154, 933 
\bibitem[1985]{sle85} Slettebak, A.\ 1985, ApJS, 59, 769 
\bibitem[1984]{tap84} Tapia M., Roth M., Costero R. \& Navarro S. 1984, Rev.\ Mex.\ Astron.\ Astrofis., 9, 65 
\bibitem[2000]{vra00} Vrancken, M., Lennon, D.\ J., Dufton, P.\ L., \& Lambert, D.\ L.\ 2000, A\&A, 358, 639 
\bibitem[1990]{wae90} Waelkens, C.\ et al.\ 1990, A\&AS, 83, 11
\bibitem[1964]{wid64} Wildey, R.\ L. 1964, ApJS, 8, 438
\bibitem[1997]{zor97} Zorec, J. \& Briot, D. 1997, A\&A, 318, 443

\end{thebibliography}
\end{document}